\documentstyle[12pt]{article}
\begin{document}
\newtheorem{algorithm}{PDP Algorithm}
\newtheorem{definition}{Definition}
\newcommand{\dia}{\mbox{diag} }
\newcommand{\tr}{\mbox{Tr}\ }
\newcommand{\aaa}{A_\alpha}
\newcommand{\ha}{H_\alpha}
\newcommand{\gab}{g_{\alpha\beta}}
\newcommand{\gba}{g_{\beta\alpha}}
\newcommand{\la}{\Lambda_\alpha}
\newcommand{\va}{V_\alpha}
\newcommand{\be}{\begin{equation}}
\newcommand{\ee}{\end{equation}}
\newcommand{\calh}{{\cal H}}
\newcommand{\ra}{\rho_\alpha}
\newcommand{\ca}{{\cal A}}
\def\baselinestretch{1.2}
\def\lqq{\lq\lq}
\def\rqq{\rq\rq}
\begin{titlepage}
%\today          \hfill
\begin{center}
\hfill    Bulletin board quant-ph/9506017 \\
\vskip .2in
{\large \bf Events and Piecewise Deterministic Dynamics
in Event-Enhanced Quantum Theory
}

\vskip .50in

Ph.~Blanchard${}^\flat$\footnote{
e-mail: blanchard@physik.uni-bielefeld.de}
\ and\ A.~Jadczyk${}^\sharp$\footnote{
e-mail: ajad@ift.uni.wroc.pl}
\vskip .2in

{\em ${}^\flat$ Faculty of Physics and BiBoS,
University of Bielefeld\\
Universit\"atstr. 25,
D-33615 Bielefeld\\
${}^\sharp$ Institute of Theoretical Physics,
University of Wroc{\l}aw\\
Pl. Maxa Borna 9,
PL-50 204 Wroc{\l}aw}
\end{center}

\vskip .2in

\begin{abstract}
We enhance the standard formalism of quantum theory to enable events.
The concepts of experiment and of measurement
are defined. Dynamics is given by Liouville's equation
that couples quantum system to a classical one. It implies a
{\sl unique}\ Markov process involving quantum
jumps, classical events and describing sample histories of
individual systems.
\end{abstract}
\end{titlepage}
\newpage
\section{Introduction}
We start with recalling John Bell's opinion on quantum
measurements.
He studied the
subject in depth and he concluded
emphasizing it repeatedly \cite{bell89,bell90}:
our difficulties with quantum measurement theory are not accidental -- they
have a reason.
He has pointed out this reason: it is
that the very concept of \lqq measurement\rqq {\em can not even be
precisely
defined within the
standard formalism}\ . We agree, and we propose a way out that has not been
tried before. Our scheme solves the essential part of the quantum
measurement puzzle - it gives a unique algorithm generating time
series of pointer readings in a continuous experiment involving
quantum systems.
We do not pretend that our solution
is the only one that solves the puzzle. But we believe that it is
a kind of a minimal solution. Even if not yet complete, it may help us to
find a way towards a more
fundamental theory.\\
The solution that we propose does not involve hidden variables
First, we point out the
reason {\em why} \lqq measurement\rqq  can not be defined within the
standard approach. That is because the standard quantum formalism has no
place for \lqq events\rqq . The only candidate
for an event that we could think of -- in the standard formalism -- is a
change of the quantum state vector. But one can not
see state vectors directly. Thus,
in order to include events, we have to extend
the standard formalism. That is what we do, and we are doing it in a minimal
way: just enough to accommodate classical events. We add explicitly
a classical part to the quantum part, and we couple classical to the
quantum.
Then we define
\lqq experiments\rqq\, and \lqq measurements\rqq\,  within the so extended
formalism.
We can show then that the standard postulates concerning measurements --
in fact, in an enhanced and refined form --
can be derived instead of being postulated.

This \lqq event enhanced
quantum theory\rqq or EEQT, as we call it,  gives experimental predictions
that are stronger than  those obtained from the standard theory. The new
theory gives answers to more  experimental questions than the old
one. It provides  algorithms for numerical simulations of
experimental time series obtained in  experiments with single quantum
systems. In particular this new theory is  falsifiable. We are working out
its new consequences for experiments, and we will report the results in due
time. But even assuming that we are successful in this respect, even then our
program will not be complete.  Our theory, in its present form, is based on
an explicit selection of an \lqq event carrying\rqq\,
 classical subsystem. But how do we select what is classical? Is it our job
or is it Nature's job? When we
want to be on a save side as much as possible, or as long as possible,
then
we tend to shift the \lqq classical\rqq\, into the observer's mind.
That was von Neumann's way out.
But if we decide to blame mind -- shall we
be save then? For how long? It seems that not too long.
This is the age of information. Soon we will need to extend our
physical theory to include a theory of mind and a theory
of knowledge. That necessity will face us anyhow, perhaps even sooner
than we are prepared to admit.  But, back to our quantum measurement
problem,  it is not
clear at all that the cut must reside that far from the
ordinary, \lqq material\rqq\ physics.
For many practical
applications the measuring  apparatus itself, or its relevant part,
can be considered classical.
We need to derive
such a splitting into classical and quantum from some clear principles.
Perhaps is is a dynamical process, perhaps the classical part is growing
with time. Perhaps time is nothing but accumulation of events. We need
new laws to describe dynamics of time itself.
At present we do not know what these laws are, we can only guess.

At the
present stage placement of the split is indeed phenomenological,
and the coupling is phenomenological too. Both are simple to handle and
easy to describe in our formalism. But where to put the Heisenberg's
cut -- that is arbitrary  to
some extent. Perhaps we need not worry too much? Perhaps relativity
of the split is a new feature that will remain with us. We do not know.
That is why we call our theory \lqq phenomenological\rqq. But we would
like to stress that the standard, orthodox, pure quantum theory is
not better in
this respect. In fact, it is much worse. It is not even able to define
what measurement is. It is not even a phenomenological theory. In fact,
strictly speaking, it is not even a theory. It is partly an art, and
that needs an artist. In this case
it needs a physicist with his human experience and with his human
intuition.
Suppose we have a problem that needs quantum theory for its solution.
Then our physicist, guided by his intuition,  will replace the
problem at hand by another problem, that can be handled. After that,
guided
by his experience, he will compute Green's function or whatsoever to
get formulas out of this other problem. Finally, guided by his previous
experience and by his intuition, he will interpret the formulas that he
got, and he will predict some numbers for the experiment.\\ That job
can not be left to
a computing machine in an unmanned space--craft. We, human beings,
may feel proud that
we are that necessary, that we can not be replaced by machines.
But would it not be better if we could spare our creativity
for inventing new theories rather than spending it unnecessarily
for application of the old ones?

In this letter paper we put stress only on the essential ideas. Details
will appear in \cite{blaja95a}, where
an extensive list of references, as well as many credits to earlier
work by other authors, are given.\\

\subsection{Summary of the results}

In this subsection we summarize the essence of our approach.
Using informal language EEQT can be described as follows:\\
Given a \lq wavy\rq\ quantum system ${\cal Q}$ we allow it to generate
distinct
classical traces - {\em events}. Quantum wave functions are not directly
observable. They may be considered as hidden variables of the theory.
On the other hand events are discrete, in principle observable directly,
real. Typically one can think of detection events
and pointer readings in quantum mechanics, but also of
creation--annihilation
events in quantum field theory. They can be observed but they do not
need an
observer for their generation (although {\em some} may be triggered by
observer's participation). They are either recorded or they
are causes for other events. It is convenient to represent events
as changes of state of a suitable  classical system. Thus formally
we divide the world into ${\cal Q}\times {\cal C}$ -- the quantum and
the classical
part. They are coupled together via a specific dynamics that
can be encoded in an irreversible Liouville evolution equation for
statistical states
of the total ${\cal Q}\times {\cal C}$ system. To avoid misunderstanding
we wish to stress it rather strongly: the fact that ${\cal Q}$ and
${\cal C}$ are coupled by a dissipative irreversible rather than by
unitary reversible  dynamics {\em does not mean}\ that
noise, or heat, or chaos, or environment, or lack of knowledge, are
involved.
In fact each of these factors, if present -- and all of them are
present in real circumstances, only blurs out
transmission of information between ${\cal Q}$ and ${\cal C}$.
The fact that ${\cal Q}$ and ${\cal C}$ must be coupled by a
dissipative rather than by reversible dynamics follows from no--go
theorems that are based on rather general assumptions
\cite{land91,ozawa92,ja94a}. We go beyond these abstract no--go
theorems that are telling us what {\em is not}\ possible.  We look
for what {\em is}\ possible, and we propose a class of couplings that,
as we believe, is optimal for the purposes of control and measurement.
With our class of couplings no more dissipation is introduced than it is
necessary for transmission of information from ${\cal Q}$ to ${\cal C}$.
Thus our Liouville equation that encodes the measurement process is to be
considered as {\em exact}\/, not as an approximate one (adding noise to
it will make it approximate). Given such
a coupling we can show that the Liouville equation encodes in a unique
way the algorithm for generating admissible histories of individual
systems. That part is new comparing with our previous paper \cite{blaja93a}.
While writing \cite{blaja93a} we did not know how to describe individual
systems. We did not suspect that for a class of couplings we are now able
to specify there is a unique event--generating algorithm.
The algorithm describes joint evolution of an individual
${\cal Q}\times {\cal C}$ system
as a piecewise deterministic process. Periods of continuous deterministic
evolution are interrupted by die tossing and random jumps that are
accompanied by changes of state of ${\cal C}$ - events.
 We call it Piecewise
Deterministic Process Algorithm, in short PDP
(the term PDP has been introduced by M.H.A. Davis -- cf. \cite{davis93}
and references therein).
The algorithm is probabilistic what reflects the fact that the
quantum world although governed by deterministic Schr\"odinger equation
is, as we know it from experience, {\em open}\/ towards the classical world
of events, and the total system ${\cal Q}\times {\cal C}$ is thus
open towards the future.  The PDP algorithm
identifies the probabilistic laws according to which
times of jumps and the events themselves are chosen.
Our generalized framework
enables us not only to gain information about the quantum system
but also to utilize it by a feed--back control of the
${\cal Q}\times {\cal C}$ coupling.
We can make the coupling dependent on the actual state of the classical
system (which may depend on the records of previous events).

Briefly, our Event--Enhanced formalism can be described as follows:
to define an {\em experiment} we must start with a division
${\cal Q}\times {\cal C}$.
Assuming, for simplicity, that ${\cal C}$ has only finite number of
states (which may be thought of as \lq pointer positions\rq\,, but they
can also represent states of a finite automaton in a quantum driven Game
of Life) $\alpha=1,\ldots ,m$, we define {\em event} as a {\em change of
state}\/
 of ${\cal C}$.
Thus there are $m^2-m$ possible events. An experiment is then described
by a specific completely positive coupling $V$ of ${\cal Q}$ and
${\cal C}$.\footnote{It
is not necessary to discuss the general concept of a completely positive
coupling here
(the interested reader can find a discussion and references in
Ref. \cite{ja94a}}
It is specified by: $(i)$ a family $H$ of quantum Hamiltonians
$H_\alpha$
parametrized by the states $\alpha$ of ${\cal C}$, $(ii)$ a family $V$ of
$m^2-m$ of quantum
operators $\gab$, with $g_{\alpha\alpha}\equiv 0$. In our previous
papers (cf. references in \cite{blaja95a})
we have described simple general rules for constructing $\gab$-s, and we
described non-trivial examples, including SQUID-tank model and
generalized \lq cloud chamber\rq\/ model that covers GRW spontaneous
localization model
as a particular, homogeneous, case.
The self--adjoint operators $H_\alpha$ determine the unitary
part of quantum evolution between jumps, while $\gab$
determine jumps, their rates and their probabilities, as well as the
non--unitary and non--linear
contribution to the continuous evolution between jumps. As an example, in
the SQUID--tank model the variable $\alpha$ is the
flux through the
coil of the classical radio--frequency oscillator circuit, and it affects,
through a transformer, the SQUID Hamiltonian. $\gab$ have also
very simple meaning there \cite{blaja93c} -- they are shifts of the
classical circuit
momentum caused by a (smoothed out, operator--valued) quantum flux.

The time evolution of
statistical states of the total ${\cal Q}\times {\cal C}$ system is
described by the Liouville equation:
\be
{\dot \rho}_\alpha=-i\left[\ha,\ra\right]+
\sum_\beta \gab\, \rho_\beta\, \gab^\star-{1\over 2}\{ \la,\ra\},
\ee
where
\be
\la=\sum_\beta \gba^\star\, \gba,
\ee
and the $\{ , \}$ stands for anti--commutator.
The operators $\ha$ and $\gab$ can be allowed to
depend explicitly on time, so that intensity of the coupling can be
controlled. Moreover, to allow for phase transitions
the quantum Hilbert space may  change with $\alpha$.
One can show that the above Liouville equation determines
a piecewise deterministic process (PDP) that generates histories
of individual systems. Within
our framework that process is {\em unique}. Our PDP is given by the
following simple
algorithm:

\begin{algorithm}
Let us assume a fixed, sufficiently small, time step $dt$.
Suppose that at time $t$ the
system is described by a quantum state vector $\psi$, $\Vert\psi\Vert=1$
and a classical state $\alpha$. Compute the scalar product
$\lambda(\psi,\alpha)=<\psi,\la\,\psi>$.
Then toss dies and choose a uniform random number $r\in [0,1]$, and jump
if $r<\lambda(\psi,\alpha)dt$, otherwise do not jump.
When jumping, toss dies and change $\alpha\rightarrow
\beta$ with probability $p_{\alpha\rightarrow\beta}=
\Vert\gba\psi\Vert^2/\lambda(\psi,\alpha)$, and change
$\psi\rightarrow\gba\psi/
\Vert\gba\psi\Vert$. If not jumping, change

$$
\psi \rightarrow
{{\exp \{-i\ha dt-{1\over2}\la dt\}\psi}\over{\Vert
\exp \{-i\ha dt-{1\over2}\la dt\}\psi\Vert}} ,\quad
t \rightarrow  t+dt.
$$ Repeat the steps.
\end{algorithm}
\noindent
{\bf Remark.}\, Another method of generating jump times is to select
a random number $r\in [0,1]$ and proceed with the continuous time evolution
by solving ${\dot \psi}= (-iH_\alpha-{1\over2}\Lambda_\alpha)\psi$
until $\Vert \psi \Vert^2=r$ - see Ref. \cite{blaja95c}
\vskip10pt
EEQT proposes that the PDP Algorithm describes {\em in an exact way}
all real events as they occur in Nature, provided
we specify correctly ${\cal Q},C,H$ and $V$.
In the following section we will formulate more precisely the basic structure
of EEQT.

\section{Mathematical scheme of  EEQT}

Let us describe mathematical framework that we use. In order to define
events, we introduce a classical system ${\cal C}$. Then possible events are
identified with changes of a (pure) state of ${\cal C}$. Let us consider
the simplest
situation corresponding to a finite set of possible events. If
necessary, we can handle infinite dimensional
generalizations of this framework. The space of states of the classical
system, denoted by ${\cal S}_c$, has $m$
states, labeled by $\alpha=1,\ldots,m$. These are the pure states of
${\cal C}$.
They correspond to possible results of single observations of ${\cal C}$.
Statistical states of ${\cal C}$ are probability measures on
${\cal S}_c$ -- in
our case just sequences $p_\alpha\geq 0, \sum_\alpha p_\alpha=1$. They
describe ensembles of observations.\\ We
will also need the algebra of (complex) observables of ${\cal C}$.
This will be
the algebra $\ca_c$ of complex functions on ${\cal S}_c$ -- in our case
just sequences $f_\alpha, \alpha =1,\ldots,m$ of complex numbers.\\ It is
convenient to use Hilbert space language even for the description of that
simple classical system. Thus we introduce an $m$-dimensional Hilbert space
$\calh_c$ with a fixed basis, and we realize $\ca_c$ as the algebra of
diagonal matrices $F=\dia(f_1,\ldots,f_m)$.\\ Statistical states of
${\cal C}$ are
then diagonal density matrices $\dia(p_1,\ldots,p_m)$, and pure states of
${\cal C}$ are vectors of the fixed basis of $\calh_c$.\\ Events are ordered
pairs of pure states $\alpha\rightarrow\beta$, $\alpha\neq\beta$. Each
event can thus be represented by an $m\times m$ matrix with $1$ at the
$(\alpha,\beta)$ entry, zero otherwise. There are $m^2-m$ possible events.
Statistical states are concerned with ensembles, while pure states and events
concern individual systems.\\ The simplest classical system is a yes--no
counter. It has only two distinct pure states. Its algebra of observables
consists of $2\times 2$ diagonal matrices.

We now come to the quantum system. Here we use the standard description.
Let ${\cal Q}$ be the quantum system whose
bounded observables are from the algebra
$\ca_q$ of bounded operators on a Hilbert space $\calh_q$.
Its pure states are unit vectors in $\calh_q$;
proportional vectors describe the same quantum state. Statistical
states of ${\cal Q}$ are given by non--negative density matrices
${\hat\rho}$,  with
$\tr ({\hat\rho})=1$. Then pure states can be identified with those density
matrices that are idempotent ${\hat\rho}^2={\hat\rho}$, i.e. with
one--dimensional orthogonal projections.

Let us now consider the total system $T=Q\times {\cal C}$. Later on we will
define \lqq experiment\rqq\, as a coupling of ${\cal C}$ to ${\cal Q}$.
That coupling will
take place within $T$. First, let us consider
statistical description, only after that we shall discuss dynamics and
coupling of the two systems. \\
For the algebra $\ca_t$ of observables of $T$
we take the tensor product of algebras of observables of ${\cal Q}$ and
${\cal C}$:
$\ca_t=\ca_q\otimes\ca_c$. It acts on the tensor product
$\calh_q\otimes\calh_c=\oplus_{\alpha=1}^m\calh_\alpha$, where
$\calh_\alpha\approx\calh_q.$ Thus
$\ca_t$ can be thought of as algebra of {\em diagonal} $m\times m$
matrices $A=(a_{\alpha\beta})$, whose entries are quantum operators:
$a_{\alpha\alpha}\in \ca_q$, $a_{\alpha\beta}=0$ for $\alpha\neq\beta$.
The classical and quantum algebras are then subalgebras of $\ca_t$;
$\ca_c$ is realized by putting $a_{\alpha\alpha}=f_\alpha I$, while
$\ca_q$ is realized by choosing $a_{\alpha\beta}=a\delta_{\alpha\beta}$.
Statistical states of ${\cal Q}\times {\cal C}$ are given by $m\times m$
diagonal matrices
$\rho=\dia(\rho_1,\ldots,\rho_m)$ whose entries are positive operators
on $\calh_q$, with the normalization $\tr (\rho)=\sum_\alpha\tr
(\ra)=1$. Tracing over ${\cal C}$ or ${\cal Q}$ produces the effective
 states of ${\cal Q}$ and ${\cal C}$ respectively:
${\hat\rho}=\sum_\alpha \ra$, $p_\alpha=\tr  (\ra )$.\\
Duality between observables and states is provided
by the expectation value $<A>_\rho=\sum_\alpha \tr (\aaa\ra)$.

We consider now dynamics. Quantum dynamics, when no information is
transferred from ${\cal Q}$ to ${\cal C}$, is described by Hamiltonians
$\ha$,
that may depend on the actual state of ${\cal C}$ (as indicated by the
index $\alpha$). They may also depend explicitly on time.  We will
use matrix notation and write $H=\dia(\ha)$.  Now take the
classical system. It is discrete here.  Thus it can not have
continuous time dynamics of its own.

Now we come to the crucial point -- the coupling.
A {\em coupling} of ${\cal Q}$ to ${\cal C}$ is specified by a matrix
$V=(\gab)$, with $g_{\alpha\alpha}=0$. To transfer  information
from ${\cal Q}$ to ${\cal C}$ we need a non--Hamiltonian term which
provides a completely positive (CP) coupling. We consider couplings
for which the evolution    equation for observables and for states is
given by the Lindblad  form: \be
{\dot A}=i[H,A]+{\cal E}\left(V^\star AV\right)-{1\over2}\{\Lambda,A\},
\ee
\be
{\dot \rho}=-i[H,\rho]+{\cal E}(V\rho V^\star)-{1\over2}\{\Lambda,\rho\},
\ee
where
${\cal E}:(A_{\alpha\beta})\mapsto \dia (A_{\alpha\alpha})$ is the
conditional expectation
onto the diagonal subalgebra given by the diagonal projection, and
\be
\Lambda={\cal E}\left(V^\star V\right).
\ee
We can also write it down in a form not involving ${\cal E}$:
\be
{\dot A}=i[H,A]+\sum_{\alpha\neq\beta}V_{[\beta\alpha]}^\star
AV_{[\beta\alpha]}-{1\over2}\{\Lambda,A\},
\ee
with $\Lambda$ given by
\be
\Lambda=\sum_{\alpha\neq\beta}V_{[\beta\alpha]}^\star V_{[\beta\alpha]},
\ee
and where $V_{[\alpha\beta]}$ denotes the matrix that has only one
non--zero entry, namely $\gab$ at the $\alpha$ row and $\beta$ column.
Expanding the matrix form we have:
\be
{\dot A}_\alpha=i[\ha,\aaa]+\sum_\beta \gba^\star
A_\beta \gba - {1\over2}\{\la,\aaa\},\label{eq:lioua}
\ee
\be
{\dot \rho}_\alpha=-i[\ha,\ra]+\sum_\beta \gab
\rho_\beta \gab^\star - {1\over2}\{\la,\ra\},\label{eq:liour}
\ee
where
\be
\la=\sum_\beta \gba^\star \gba.
\ee
Again, the operators $\gab$ can be allowed to depend explicitly
on time.

Following \cite{ja94c} we now define {\em experiment} and
{\em measurement}:

\begin{definition}
An {\bf experiment} is a CP coupling between a quantum and a classical
system. One observes then the classical system and attempts to learn
from it about characteristics of state and of dynamics of the quantum
system.
\end{definition}

\begin{definition}
A {\bf measurement} is an experiment that is used for a particular
purpose: for determining values, or statistical distribution of values,
of given physical quantities.
\end{definition}

\noindent{\bf Remark.}\,
The definition of experiment above is concerned with the {\em conditions}
that define it. In the next sections we will discuss the PDP
algorithm that simulates a typical {\em run} of a given experiment. In
practical situations it is rather easy to decide what constitutes
${\cal Q}$,
what constitutes ${\cal C}$ and how to write down the coupling. Then, if
necessary, ${\cal Q}$ is enlarged, and ${\cal C}$ is shifted towards more
macroscopic
and/or more classical. However the new point of view that we propose
allows us to consider our whole Universe as \lq experiment\rq\,
and we are witnesses and participants of one particular run. Then the
question arises: {\em what is the true} ${\cal C}$? This question is yet
to be answered. Some hints can be found in the closing section of
Ref. \cite{blaja95a}.

\section{Statistical ensembles, individual systems,
and the PDP algorithm}
Time evolution in the standard quantum theory of closed systems is unitary
reversible. In quantum theory of open systems, dissipative,
irreversible
evolution is being used. But there it is considered only as an approximate
description, not the exact one. It is useful
when external unknown factors disturb the true unitary dynamics, and we
either do not need, or are not able because of computational complexity,
to use the exact unitary dynamics. The main
difference between unitary reversible and dissipative irreversible evolutions
is in their mixing properties. Unitary
evolution
maps pure states into pure states, while dissipative one maps pure states
into mixtures. Pure states describe individual systems. Mixtures describe
statistical ensembles. Thus when evolution preserves purity of states, then
we may assume that it concerns individual systems.
Things change when we want to move from
mere lasting to events that happen in time.
{}From continuous and deterministic evolution of possibilities to discrete
realization of actualities, when God allows us either to rely on chance or
to choose. Standard quantum theory is helpless when it comes
to generation of events. But the material world around us, the living nature,
the phenomena that we want to understand  -- all that -- we perceive only
through events, and nothing but events. Thus standard quantum
theory must be enhanced. The only way to make quantum system to be coupled to
a classical event--carrying system is via dissipative dynamics as described
in the previous section. But the Liouville equation with a nontrivial
coupling term must lead from pure states to mixed states. Thus it
does not describe individual systems - it describes statistical ensembles.
What describes individual systems is the PDP Algorithm as given at the end
of Sec.1.1. A priori one could think that there may be many such algorithms
with the property that, after averaging over individual sample paths,
reproduce a given statistical
behaviour. Here that is not so. We have shown that PDP Algorithm is unique.
The proof is given in an infinitesimal form  in \cite{blaja95a}.
A rigorous global proof can be found in \cite{jakol}. This fact, i.e.
uniqueness of the random process that reproduces master equation,
distinguishes PDP from Quantum Monte Carlo methods used in quantum optics.
We are discussing this fact in some details in \cite{blaja95a}.

The PDP Algorithm is the most important new result of our approach. It
is simple, it is universal, it is useful. We have already mentioned it
in the introduction that all the standard postulates of quantum theory
about measurements and their probabilities can be deduced from the PDP
via suitable couplings. We have discussed this subject elsewhere in the
aforementioned references. In particular we succeeded in reproducing
real time formation of particle tracks and interference patterns
\cite{ja94b}. We are investigating new applications of the algorithm.
But for a successful applications in new situations we need one more
piece in the theory - a piece that is still missing. We know how
to describe {\em measurements}, but we must also know how to describe
state preparations. In principle state preparation can be thought of as
a measurement with sample selection, so it could essentially fit into
the scheme that we have already described. However, we need more. We
need to learn how to describe preparation of multiparticle states that
look like individual particle states at each given time. Only then we
will be able to make realistic simulation of experiments in neutron
interferometry or electron holography, when a source produces weak but
coherent particle beams. Work in this direction is in progress.

\vskip10pt
\noindent
{\bf Acknowledgements}\\
One of us (A.J) acknowledges support of A. von Humboldt
Foundation extended during various periods of work on this paper.

\end{document}